\documentclass{article}

\usepackage{arxiv}
\usepackage{amsmath}
\usepackage[utf8]{inputenc} 
\usepackage[T1]{fontenc}    
\usepackage{hyperref}       
\usepackage{url}            
\usepackage{booktabs}       
\usepackage{amsfonts}       
\usepackage{nicefrac}       
\usepackage{microtype}      
\usepackage{lipsum}
\usepackage{graphicx}
\graphicspath{ {./images/} }

\title{Investigating the Violation of Charge Parity Symmetry Through Top Quark Chromo-Electric Dipole Moments by Using Machine Learning Techniques}

\author{
 Bora Işıldak \\
  Physics Department\\
  Yıldız Technical Univesity\\
    \texttt{bora.isildak@yildiz.edu.tr} \\
   \And
 Alper Hayreter\\
  Mathematical and Natural Sciences\\
  Ozyegin university\\
   \texttt{alper.hayreter@ozyegin.edu.tr} \\
  \And
 Murat Hüdaverdi \\
  Physics Department \\
  Yıldız Technical University\\
    \texttt{hudaverd@yildiz.edu.tr} \\
  \AND
  Fatih Ilgın \\
  Mathematical and Natural Sciences\\
  Ozyegin university\\
  \texttt{fatih.ilgin@ozyegin.edu.tr} \\
  \And
  Sinem Salva\\
  Mathematical and Natural Sciences\\
  Ozyegin university\\
  \texttt{sinem.salva@ozyegin.edu.tr} \\
  \And
  Ebru Şimşek \\
Mathematical and Natural Sciences\\
  Ozyegin university\\
  \texttt{ebru.simsek@ozyegin.edu.tr} \\
  \And
  Sinan Güyer \\
  Physics Department\\
  Marmara university\\
  \texttt{sinanguyer@gmail.com} \\
}

\DeclareUnicodeCharacter{2212}{-}

\begin{document}
\maketitle
\begin{abstract}
There are a number of studies in the literature on search for Charge-Parity (CP) violating signals in top quark productions at the LHC. In most of these studies, ChromoMagnetic Dipole Moments (CMDM) and ChromoElectric Dipole Moments (CEDM) of top quarks is bounded either by deviations from the Standard Model (SM) cross sections or by T-odd asymmetries in di-muon channels. However, the required precision on these cross section values are far beyond from that of ATLAS or CMS experiments can reach. In this letter, the investigation of CEDM based asymmetries in the semileptonic top pair decays
are presented as T-odd asymmetries in CMS experiment. Expected asymmetry values are determined at the detector level using MadGraph5, Pythia8 and Delphes softwares along with the discrimination of the signal and the background with Deep Neural Networks (DNN). 
\end{abstract}


\section{Introduction}
\label{sec:introduction}
There are many unsolved problems in the Standard Model (SM) of particle physics. One of those interesting problems is the baryon asymmetry that is being observed in the current state of the universe. CP violation is a necessary condition to explain the asymmetry according to Sakharov conditions \cite{Sakharov:1967dj}, and is allowed in electroweak theory only via complex phases in the Cabbibo-Kobayashi-Maskawa (CKM) matrix \cite{Cabbibo, Kobayashi-Maskawa}. However, the amount of CP violation due to CKM complex phases is not  enough to match with the observed asymmetry. Therefore, new sources of CP violation beyond the SM should be expected. The top quark is a prefect candidate to search for evidences of those new sources due to its unique properties among quarks, such as its high mass, small lifetime ($\tau=5\times10^{-25}$ s). Moreover, unlike the other quarks, top quark does not hadronize but decays through dileptonic, semileptonic or fully hadronic channels. These decay channels provide a prefect laboratory for searching new physics effects.  

The CERN Large Hadron Collider (LHC) practically runs as a top quark factory. Hence, it provides a wide spectrum of opportunities to search for new sources of CP violation at high energies. The simplest phenomenological approach is to study the lowest-dimension operators that can produce the desired CP violation. Since the 90\% of the top quarks are produced via gluon fusion at the LHC \cite{Heinemeyer:2013tqa}, the top-quark CEDM becomes a very crucial phenomenon for this type of studies. A non-vanishing CEDM is an indication of time-reversal violation or equivalently a CP violation because of the CPT theorem. Due to the fact that the CEDM arises at three loops in SM  and its value has been estimated to be negligibly small ($\leq 10^{-30}$g$_{s}$.cm) \cite{Czarnecki:1997bu}, an observation of a considerable CEDM would be a strong sign for new physics beyond SM. 

Top quarks magnetic and electric dipole couplings to gluons conventionally written as the following dimension-five effective Lagrangian

\begin{equation}
\mathcal{L}=\frac{g_{s}}{2} \bar{t} T^{a} \sigma^{\mu \nu}\left(a_t^g+i \gamma_{5} d_t^g\right) t G_{\mu \nu}^{a},
    \label{Eq1}
\end{equation}
where $G^{a}_{\mu \nu}$ is the gluon stress tensor and $T^a$ are the $SU(3)$ generators. $a_t^g$ and $d_t^g$ are called as CMDM and CEDM respectively. The gauge invariant generalization of Equation (\ref{Eq1}) can be given as

\begin{equation}
    \mathcal{L}=g_{s} \frac{d_{tG}}{\Lambda^{2}} \bar{q}_{3} \sigma^{\mu \nu} T^{a} t \tilde{\phi} G_{\mu \nu}^{a}+\mathrm{H.c.}
    \label{Eq2}
\end{equation}
\par
 where $q_{3}$ is the third generation SM quark doublet, $\phi$ is the SM Higgs doublet, $\tilde{\phi}_{i}=\epsilon_{ij}\phi_{j}$ and Tr$(T_{a}T_{b})=\delta_{ab}/2$. When the Higgs field acquires a vacuum expectation value we have the following correspondence, $d_{t}^{g}=(\sqrt{2}v / \Lambda^{2})\  \text{Im}(d_{tG})$. In this study, the new physics scale $\Lambda$ is imposed as 1 TeV. However, a different scale can be easily inferred from Equation \ref{Eq2}.

There are plenty of top quark CMDM and CEDM studies in the literature both within the context of SM and with several extensions of SM \cite{Bernreuther:1992be,Brandenburg:1992be,Atwood:1992vj,Bernreuther:1993hq,Cheung:1995nt,Choi:1997ie,Sjolin:2003ah,Martinez:2007qf,Antipin:2008zx,Gupta:2009wu,Gupta:2009eq,Choudhury:2009wd,Hioki:2009hm,HIOKI:2011xx,Kamenik:2011dk,Ibrahim:2011im,Hioki:2012vn,Biswal:2012dr,Baumgart:2012ay,Hioki:2013hva,Bernreuther:2013aga,Kiers:2014uqa,Englert:2014oea,Rindani:2015vya,Gaitan:2015aia,Bernreuther:2015yna}. In recent studies, dileptonic and semileptonic decays of top quark pairs were discussed, and considering ATLAS and CMS results certain bounds on CMDM and CEDM were obtained with the help of deviations from $\sigma_{t\bar{t}}^{\rm{SM}}$ as well as the asymmetries on several triple product operators in decay channels \cite{PhysRevD.88.034033, Hayreter2016}. However, those studies consider the parton level objects for calculating the asymmetries and treating the background physics processes as dilution factors in the asymmetry. In this study, the asymmetries were calculated at the detector level by using simulated event samples for along with the major background contributions. Moreover, using event variables, a DNN model were trained to eliminate the background contribution resulting a better sensitivity on the expected asymmetry.

The paper is organized as follows. In Sec. \ref{sec:event_simulation}, simulation details of the signal and background events are given. Reconstruction of the events and pre-selection criteria applied to the events are described in Sec. \ref{sec:event_selection}. Sec. \ref{sec:signal_background_discrimination} is devoted to the discussion of signal and the background discrimination via DNN model. Asymmetries calculated after all the selection processes applied on both signal and background events are reported in Sec. \ref{sec:asymmetries}. Finally, a conclusion is given in Sec. \ref{sec:results_and_conclusion}.

 \begin{figure}[h!]
 \begin{center}
\includegraphics[scale=0.5]{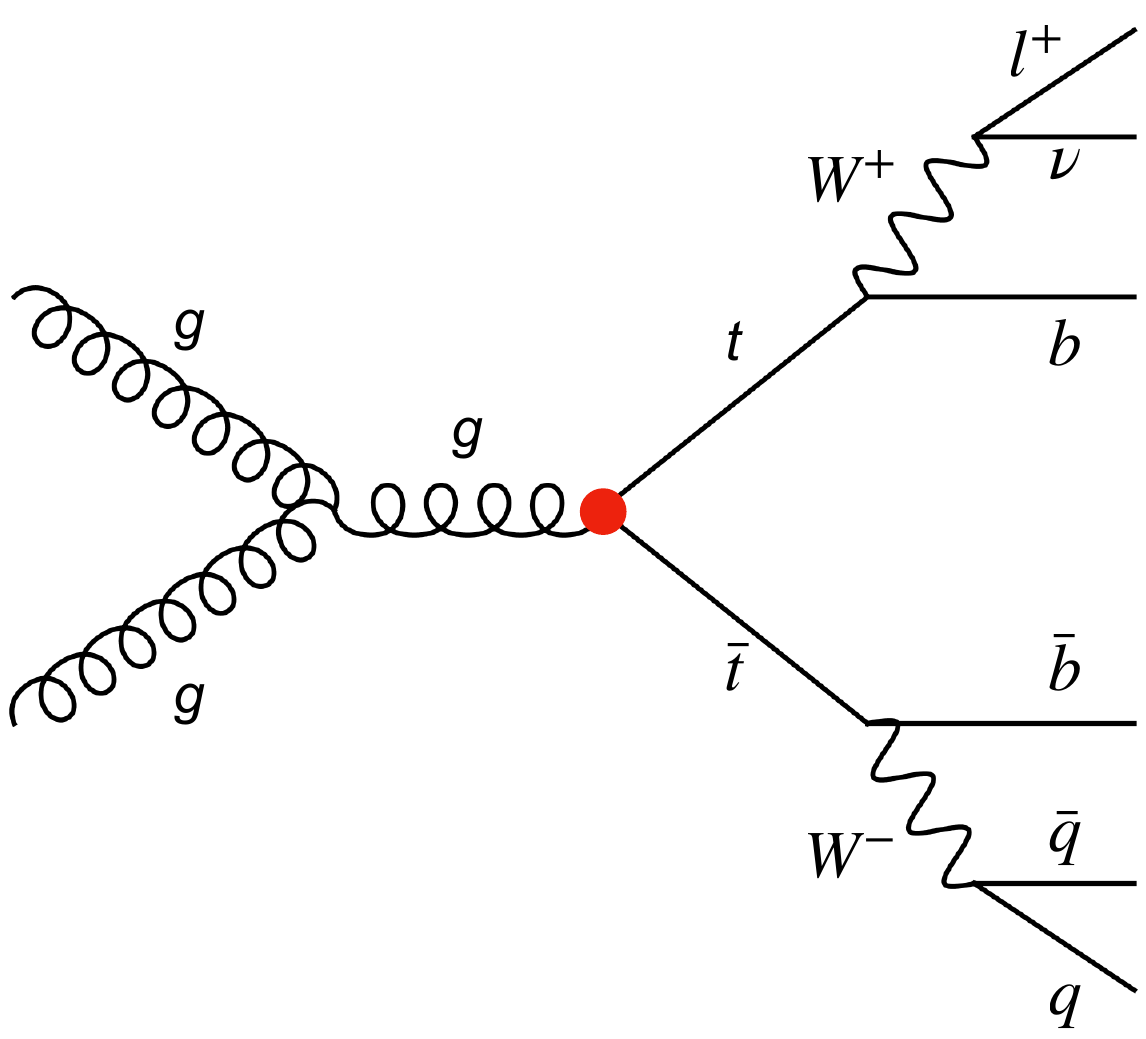}
\caption{Feynman diagram for the top quark pair semileptonic decay. The red dot represents the CEDM contribution.}
\label{fig:signal_feyn_diag}
\end{center}
\end{figure}   

\section{Event Simulation}
\label{sec:event_simulation}
In order to understand the possibility of measuring the asymmetries in the triple-product distributions which are going to be discussed in Sec. \ref{sec:asymmetries}, multiple event samples were generated  at 14 TeV of center-of-mass energy for the process $pp \rightarrow t\bar{t} \rightarrow b \bar{b} \ell^{\pm} \nu j j$ with $\ell$=$\mu$,$e$. The CEDM coupling is implemented exactly as in \cite{Hayreter2016} using $\textsc{MadGraph5}$ \cite{MadGraph} along with the $\textsc{FeynRules}$ \cite{FeynRules} package. Resulting UFO files were used to generate signal samples with different values of the coupling parameter $d_t^g$ described in Equation \ref{Eq2}. Parton shower and hadronization processes are simulated using $\textsc{Pythia8}$ \cite{SJOSTRAND2008852} with MLM matching scheme \cite{Mangano_2007}. Finally, detector level data were obtained using a fast detector simulation software $\textsc{Delphes}$ \cite{Delphes} with default CMS detector specification card with no \nolinebreak{pile-up} that comes with the $\textsc{MadGraph5}$ version 2.6.7.

W plus jets, Drell-Yan plus jets and single-top (W plus top and t-channel) processes are considered as background since they may produce the semileptonic $t\bar{t}$ decay signature. As a comparison, SM top pair production cross section is also included in Table \ref{table:data}.  None of these background processes is CP violating in the SM. However, their contribution can considerably decrease the observed asymmetry values as a result of the larger measured cross section of the $t\bar{t}$ events.  QCD multijet background was omitted since the requirement of an isolated lepton in the event  strongly filters this background. Table \ref{table:data} shows the cross-sections of those background processes. W plus Jets and Drell-Yan plus Jets samples were generated using \textsc{MadGraph5} followed by parton shower and hadronization simulations via \textsc{Pythia8}. These events were also simulated at the detector level using \textsc{Delphes}. Single-top events were generated with \textsc{Powheg-Box} \cite{Re:2010bp, Alioli:2010xd, Nason:2004rx, Frixione:2007vw, Alioli:2010xd}. 

\begin{table}[ht]
\centering
\begin{tabular}{lc}
 Background Process & Cross-section [pb]  \\ [0.5ex] 
 \hline\hline
 W+Jets & 61527 \\ 
 \hline
 Drell-Yan + Jets & 5765  \\
 \hline
 Single-top (Wtop) & 71.7  \\
 \hline
 Single-top (t-channel) & 219.6  \\  
\hline \hline
 Top pair production & 985.7  \\  
\hline
\end{tabular}
\caption{Background cross sections.}
\label{table:data}
\end{table}

Drell-Yan and W$^{\pm}$ boson samples were generated by MADGRAPH allowing up to three associated jets with $p_T>$ 20 GeV, and a sample of only those with semileptonic decay signature were selected. 

\section{Event Reconstruction and Selection}
\label{sec:event_selection}
The semileptonic decay topology of the top quark pairs requires at least two b-tagged jets along with the two light flavor jets (u,d,c or s quark jets). Moreover, the signal event must have an isolated lepton and corresponding neutrino from the W$^{\pm}$ boson decay. Since the neutrinos are not directly detectable, there should be a substantial imbalance in the transverse energy quantified as the missing energy transverse (MET). For the lepton selection an isolation criteria were used. These criteria are defined such that the $p_T$ sum of all stable particles, excluding the leptons and neutrinos, within $\Delta R$ = 0.4 should be less than 12\% (25\%) of the electron (muon) $p_T$.

\begin{table}[h!]
\centering
\renewcommand{\arraystretch}{1.5}
\begin{tabular}{cc}
  Required Object & Requirement\\
 \hline \hline
 $e(\mu)$~~~~~ &  {$p_T>$ 20 GeV, $|\eta|<$ 2.4} \\
 & {Iso $<0.12$ ($0.25$)}\\
 \hline 
 at least 2 b-tagged jets ~~~~~& $p_T>$ 25 GeV, $|\eta|<$ 2.4 \\
 \hline
  at least 2 light flavour jets ~~~~~& $p_T>$ 20 GeV, $|\eta|<$ 2.4 \\
 \hline
 missing transverse energy ~~~~~& MET$>30 $ GeV \\
 \hline
\end{tabular}
\caption{Event selection criteria.}
\label{table:event_selection}
\end{table}

Events with exactly one electron or with exactly one muon passing all selection criteria given in Table \ref{table:event_selection} are selected. In addition to the lepton, at least 4 jets are required where two of them must be b-tagged with $p_T>$ 25 GeV  and the other two must be light flavored with $p_T>$ 20 GeV. As a final requirement, all events must have a MET value greater than 30 GeV. After the selection requirements were applied, event variables to be used in the DNN training were calculated and saved to a file for the training and the asymmetry analysis. Triple products which will be mentioned in Section \ref{sec:asymmetries} were also calculated and recorded at this step.

\section{Signal and Background Discrimination}
\label{sec:signal_background_discrimination} 
The event selection criteria described in Section \ref{sec:event_selection} eliminate most of the background events coming from different SM physics processes. However, the number of remaining background events after this selection may still have a substantial dilution effect on the asymmetry (see Table \ref{table:cut_flow}). Therefore, a second selection step was developed by training a Machine Learning (ML) model to further discriminate the signal events from the background events. First, a SM  sample for $pp \rightarrow t\bar{t} \rightarrow b \bar{b} \ell^{\pm} \nu j j$  with $\ell=\mu,e$ has been  generated to be used in the training. The training was performed by using several event and physics object variables as the inputs of the model. Event variables such as the number of jets, number of b-tagged jets, MET, azimuthal angle  of the missing transverse energy (MET-$\phi$), $p_T$ sum of all jets in the event (scalar HT), sphericity, aplanarity  and the first four Fox-Wolfram moments were used. Definition of these variables can be found elsewhere \cite{event_variables}. Additionally, physics object variables such as the transverse momenta of the leading four jets and transverse momentum of the selected lepton were used in the input vector. As one might expect, there remained more events in the signal sample than those remained in each different background samples after applying the object selection criteria. For a balanced learning process, each of the background samples have been up-sampled  to match their sizes with the signal sample size. All input values were standardized to zero mean and unit variance before the training. The sample were split into training and validation sub-samples with 1:1 ratio. 
\begin{figure}[ht]
\centering
  \includegraphics[width=\linewidth]{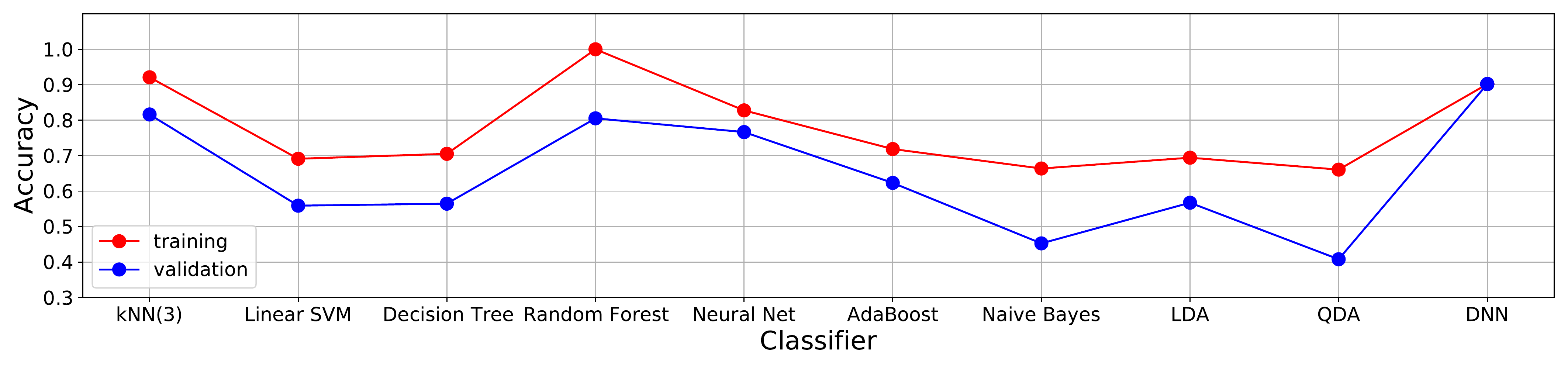}
  \caption{Performance comparison of different classifiers for training and validation samples.}
  \label{fig:model_comparison}
\end{figure}

For choosing the appropriate ML algorithm, several traditional ML classifiers were compared along with the DNN classifier in terms of accuracy and overfitting of the model. In Figure \ref{fig:model_comparison}, it can be seen that the DNN is the best choice among the algorithms compared. The only algorithm that give a training accuracy better than the DNN classifier are the K-Nearest Neighbors (kNN) with 3 nearest neighbours and the Random Forest . However, they achieve much worse on the validation sample which is a clear indication that these classifiers overfit the training sample.

\begin{figure}[ht]
\centering
  \includegraphics[width=\linewidth]{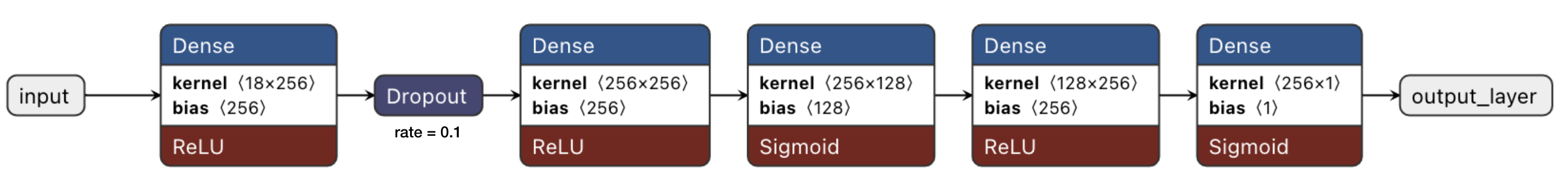}
  \caption{Layer structure of DNN trained for the signal-background discrimination.}
  \label{fig:model_diagram}
\end{figure}

\begin{figure}[ht]
\centering
  \includegraphics[width=0.6 \linewidth]{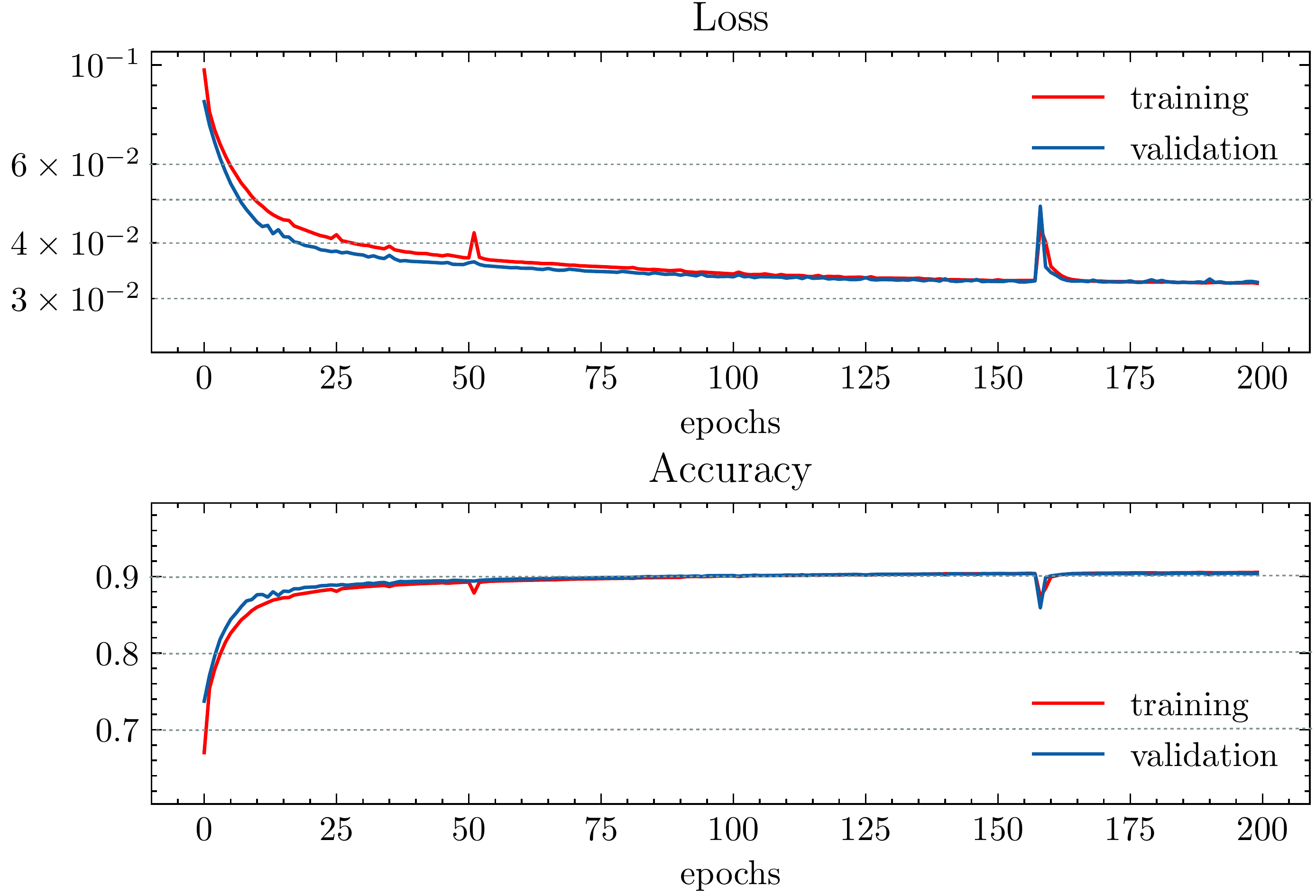}
  \caption{Evolution of the loss values (top) and the obtained accuracies (bottom) for both training and validation samples.}
  \label{fig:loss_acc_vs_epoch}
\end{figure}

\begin{figure}[ht]
\centering
  \includegraphics[width=0.45 \linewidth]{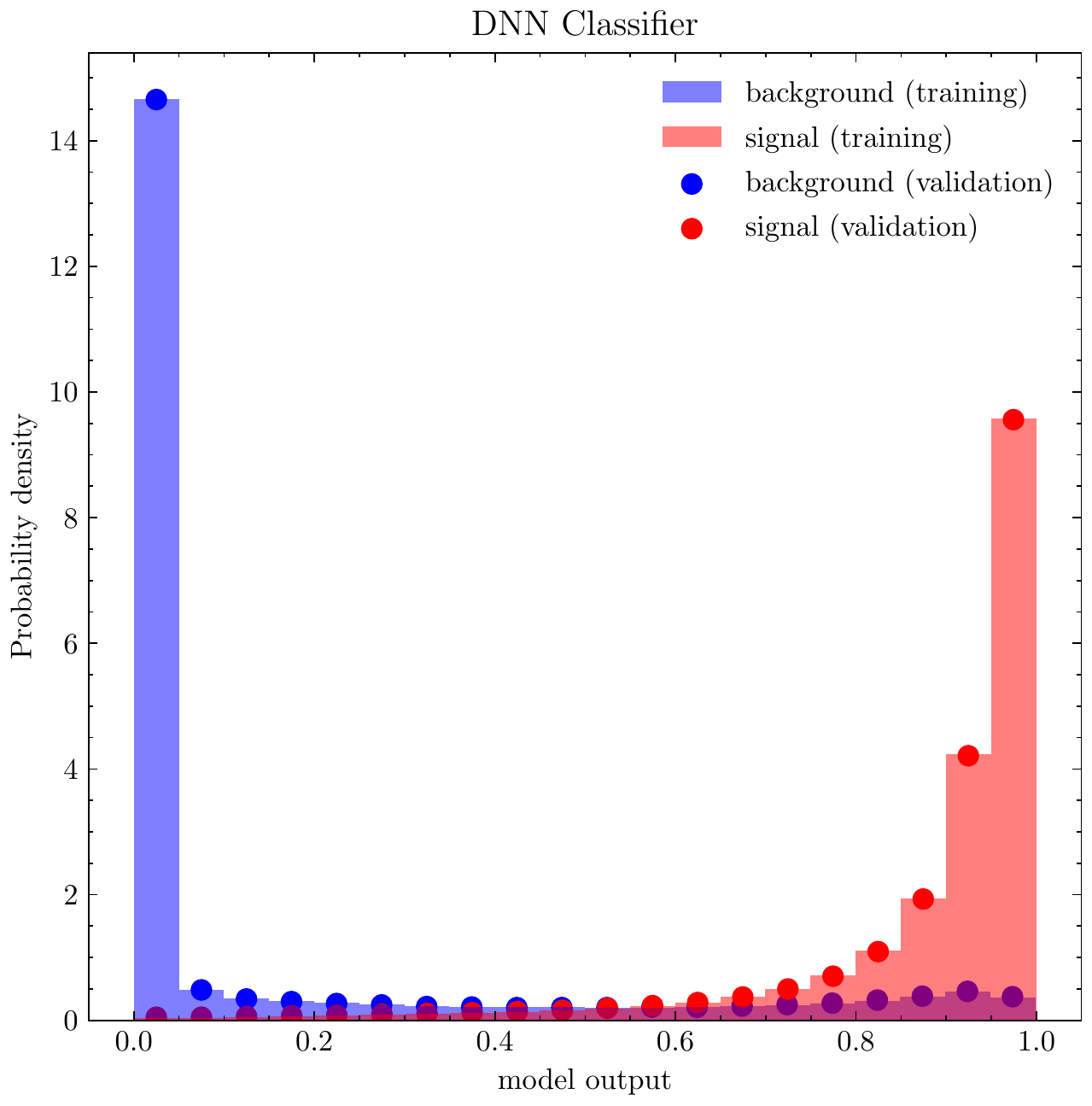}
  \includegraphics[width=0.45 \linewidth]{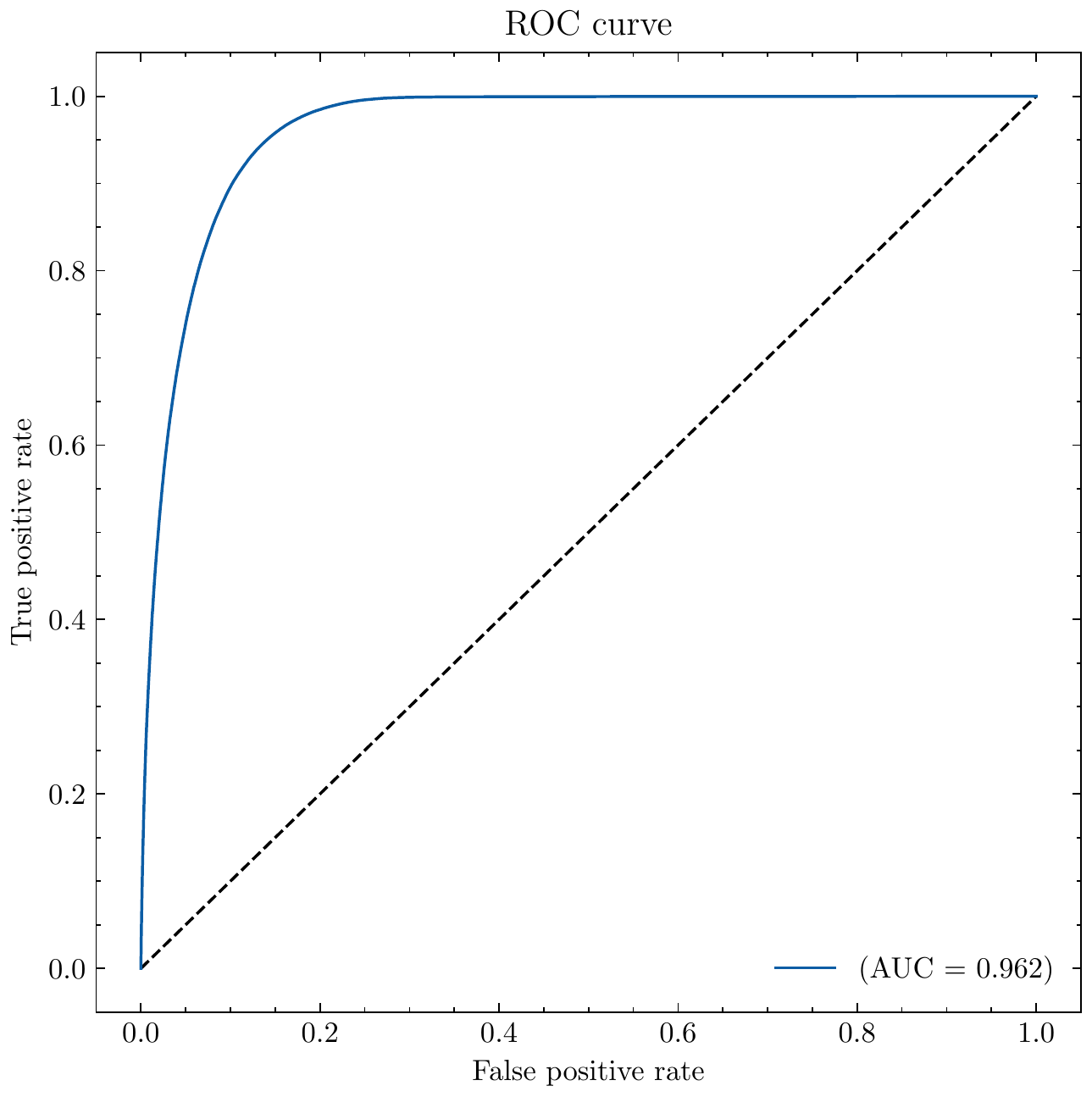}
  \caption{Left panel: DNN output distribution for the background events in the training sample (blue shaded), for the background events in the test sample (blue dots), for the signal events in the training sample (red shaded) and for the signal events in the test sample (red dots).  Right Panel: Receiver Operator Characteristics (ROC) curve for the trained model.}
  \label{fig:model_output_dist}
\end{figure}

In order to construct the optimum DNN structure, a hyper-parameter search was performed using a grid search algorithm for a model with four fully connected hidden layers.  All 1944 different parameter configurations (see Table \ref{table:hpo search}) were separately trained for 50 epochs to find the set with the highest validation accuracy.  The result of the hyper-parameter search suggested to use the network structure  given in Figure \ref{fig:model_diagram}.  Then the model with the optimum hyper-parameters has been trained with Adam optimiser (learning rate = 0.001) \cite{Adam} for 200 epochs to obtain the final classifier model to be used in the asymmetry  calculation. The top panel of Figure \ref{fig:loss_acc_vs_epoch} shows the evolution of the loss values during epochs for both training and validation samples. In the bottom panel, accuracies can be seen again for both training and validation samples. The classification accuracy for both training and the test set reaches to plateau at 90.4\%. Although the network used is relatively small, it is necessary to check whether it overfits on the training sample. One way of investigating the model overfit is looking at the model output distributions for individual class labels separately on the training and test samples. If the distributions for a given class label show a significant discrepancy between training and test samples, this discrepancy can be considered as a clear indication of overfitting model. In the left panel of Figure \ref{fig:model_output_dist},  the normalized DNN model output distributions for both signal and background events separately on the training and validation samples are shown. No significant difference has been observed between the distributions obtained from the training and the validation samples for both signal  and background events.  The Receiver Operator Characteristics (ROC) curve is also given in the right panel of Figure \ref{fig:model_output_dist}  where the Area Under Curve (AUC) score is found to be 0.962. The number of events for each generated background sample and the remaining events after event selection and DNN selection steps are given in Table \ref{table:cut_flow}.

\begin{table}[!h]
\renewcommand{\arraystretch}{1.5}
\begin{tabular}{ll}
 \multicolumn{1}{c}{Parameter Name}~~~~~ &  \multicolumn{1}{c}{Parameter Values}\\
\hline\hline
Hidden Layer 1 Activation~~~~~ & {[}relu, tanh, sigmoid{]}$\times${[}256{]} \\ \hline
Dropout Layer (Dropout Rate)~~~~~  & {0, 0.1} \\ \hline
Hidden Layer 2 Activation~~~~~ & {[}relu, tanh, sigmoid{]}$\times${[}256{]} \\ \hline
Hidden Layer 3 Activation~~~~~ & {[}relu, tanh, sigmoid{]}$\times${[}128, 256{]} \\ \hline
Hidden Layer 4 Activation~~~~~ & {[}relu, tanh, sigmoid{]}$\times${[}256{]} \\ \hline
Last Layer Activation~~~~~     & {[}sigmoid, tanh, softmax{]}                            \\ \hline
Loss Function~~~~~ & {[}logcosh, mse{]}         \\ \hline
\end{tabular}
\caption{Hyper-parameters used for the grid search. Note that the Cartesian product is represented by ``$\times$'' symbol.}
\label{table:hpo search}
\end{table}

\begin{table}
\centering
\begin{tabular}{cccc}
& Generated  & Event Selection & DNN Selection\\
  \hline \hline
  W+Jets   & 6.0e+07  & 608 & 58\\
  \hline
  Drell-Yan + Jets  & 3.0e+07 & 918 & 88\\
  \hline
   Single-top & 1.0e+06  & 841751 & 80808\\
   \hline
\end{tabular}
\caption{Number of remaining events after each selection step for generated background samples.}
\label{table:cut_flow}
\end{table}

\section{ Triple Product Asymmetries}
\label{sec:asymmetries}
Since weak interactions violate parity, collider processes that can involve weak interactions typically show asymmetry in the distributions of final-state particles. These asymmetries are typically sensitive to the difference in interactions between particles and antiparticles. Therefore, they can be used to distinguish a small asymmetric signal from a large but symmetrical background as a precise measure of differences in interaction strengths. CP violating couplings can be investigated via the triple products of four momenta of proton beam and final state particles that involve $t$ quark, $b$ quark, lepton and jet momenta. Further details of the triple product definition can be found in \cite{Hayreter2016} where 14 different triple product operators were used. Since distinguishing $b$ quarks from $\bar{b}$ quarks experimentally is not an easy task, among those 14 operators, even ones under the interchange of $b$ and $\bar{b}$ quarks were considered. Also, operators requiring $t$ quark reconstruction were excluded. As a result, the asymmetries for the following five operators were calculated by using Equation \ref{eqn:asymmetry definition}. 

\begin{eqnarray}
\begin{aligned}
    & \mathcal{O}_1 = \epsilon\left(b^{\ell}, b^{j}, \ell, j\right), \\
    & \mathcal{O}_2 = (q \cdot \ell) \epsilon(b+\bar{b}, q, \ell, j), \\
    & \mathcal{O}_3 = (q \cdot (b-\bar{b})) \epsilon(b, \bar{b}, q, j), \\
    & \mathcal{O}_4 = (q \cdot (b-\bar{b})) \epsilon(P, q, b, \bar{b}), \\ 
    & \mathcal{O}_5 = q_\ell \epsilon(P, b+\bar{b}, \ell, j), 
\end{aligned}
\label{eqn:operators}
\end{eqnarray}

\noindent where $q^{\mu}=p_{1}^{\mu}-p_{2}^{\mu}$ the difference of beam four-momenta, $P^\mu=p_{1}^{\mu}+p_{2}^{\mu}$ the sum of beam four-momenta. It should be noted that the dot products in these equations are 4-vector dot products.  In triple product $\mathcal{O}_1$, $b$ quarks are labeled as $b^\ell$ and $b^j$ depending on their proximity to the lepton. More explicitly, among two $b$ quark jets with the highest $p_T$, the one that is closer to the lepton is labeled as $b_\ell$ and the other one is labeled as $b_j$.

The $\epsilon(p_1, p_2, p_3, p_4)$ function is defined as,

\begin{equation}
\epsilon(p_1, p_2, p_3, p_4)=\epsilon_{\mu \nu \alpha \beta} p_{1}^{\mu} p_{2}^{\nu} p_{3}^{\alpha} p_{4}^{\beta}
 \label{eqn:epsilon definition}
\end{equation}

 with $\text { Levi-Cività}$ tensor $\epsilon_{0123}=-1$. Equation \ref{eqn:epsilon definition} also be written as a determinant, 

\begin{equation}
\epsilon(p_1, p_2, p_3, p_4)=\left|\begin{array}{cccc}
p_{1 x} & p_{1 y} & p_{1 z} & E_{1} \\
p_{2 x} & p_{2 y} & p_{2 z} & E_{2} \\
p_{3 x} & p_{3 y} & p_{3 z} & E_{3} \\
p_{4 x} & p_{4 y} & p_{4 z} & E_{4}
\end{array}\right|
 \label{eqn:epsilon definition determinant}
\end{equation}

For a given triple product $\mathcal{O}_i$, the asymmetry in the laboratory frame is defined as

\begin{equation}
    \mathcal{A}_i =\frac{\sigma\left(\mathcal{O}_i>0\right)-\sigma\left(\mathcal{O}_i<0\right)}{\sigma_{\text{SM}}}.
    \label{eqn:asymmetry definition}
\end{equation}

\section{Results and Conclusion}
\label{sec:results_and_conclusion}
\begin{figure*}[!h]
\centering
  \includegraphics[width=0.87 \linewidth]{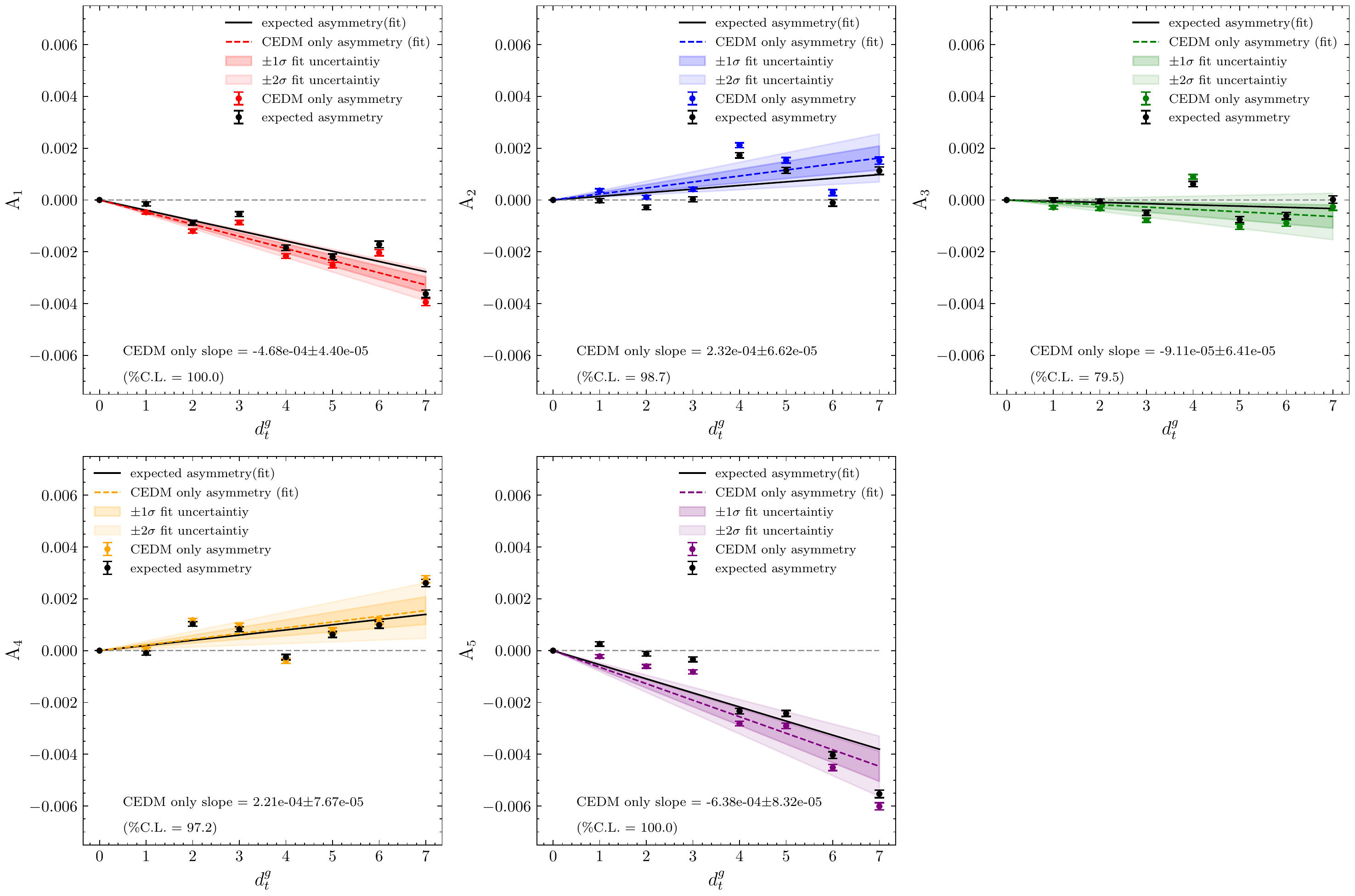}
  \caption{Asymmetries obtained for different triple products denoted by $A_i$ without DNN signal classification with CMS fast detector simulation for 3000 fb$^{-1}$ of integrated luminosity.}
  \label{fig:asymmetries_wo_DNN}
\end{figure*}

\begin{figure*}[!h]
\centering
  \includegraphics[width=0.87 \linewidth]{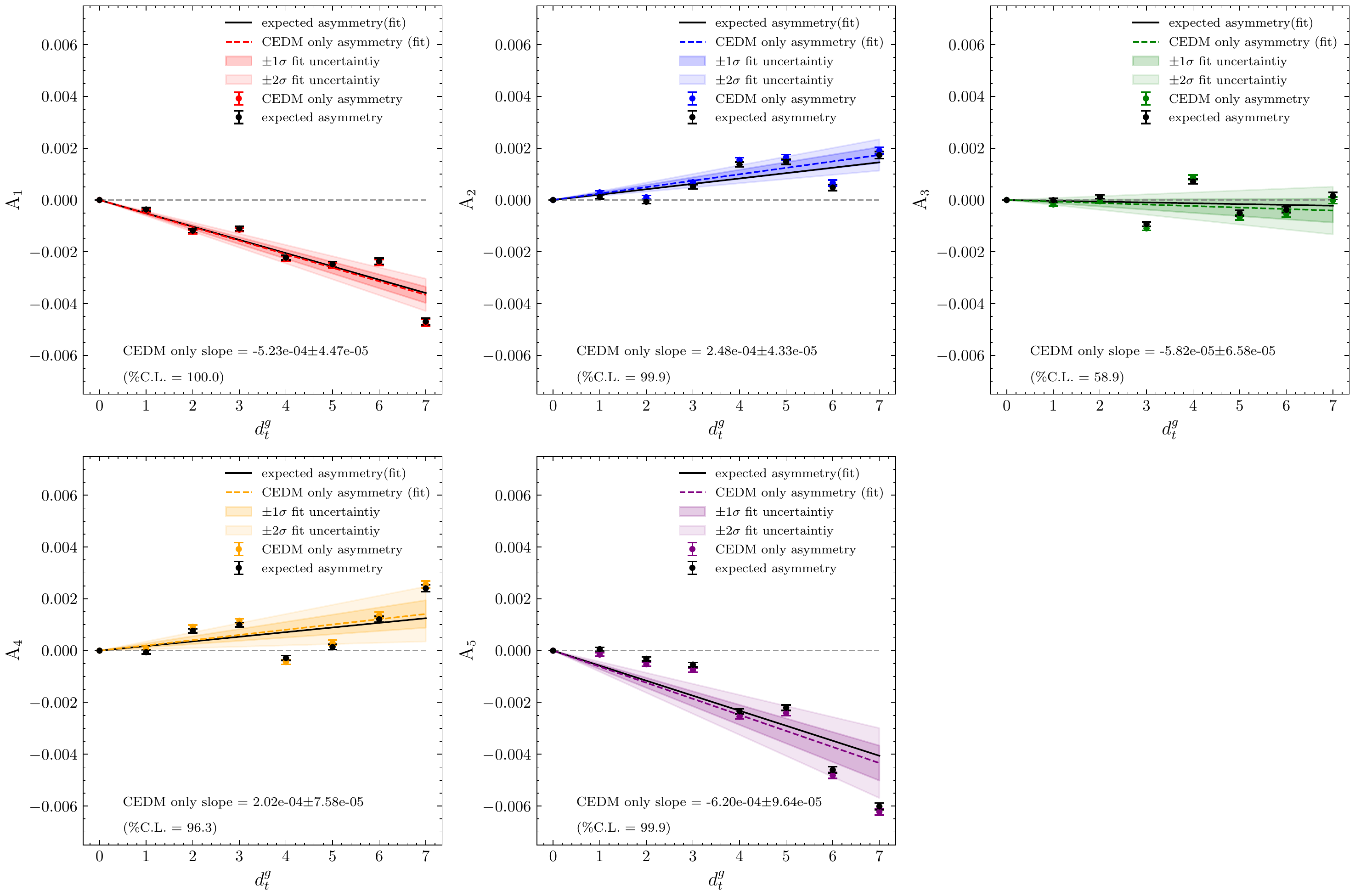}
  \caption{Asymmetries obtained for different triple products denoted by $A_i$ with DNN signal classification with CMS fast detector simulation for 3000 fb$^{-1}$ of integrated luminosity.}
  \label{fig:asymmetries_DNN}
\end{figure*}

Starting from the Lagrangian in Equation \ref{Eq2}, which includes new physics contribution for semileptonic $t\bar{t}$ decays, detector level signal events were simulated for integer $d_{tG}$ values starting from 1 to 7. A recent study from CMS collaboration \cite{CMS:2022voq} puts a limit on the $d_{tG}$ value an order of magnitude smaller because they measured insignificant asymmetries. Since our purpose is to prove the usefulness of extra DNN selection procedure, we boosted $d_{tG}$ values to have significant asymmetries. $d_{tG}=0$ sample is the generated background events and cross section values are given in Table \ref{table:data}. Event selection criteria used for eliminating the SM background processes were shown in Table \ref{table:event_selection}. In addition, a DNN classifier is trained for discriminating the signal events from the background events in order to suppress the dilution effect of the background contribution in the asymmetries. Asymmetries are calculated for five different operators given in Equation \ref{eqn:operators} as a function of $d_{tG}$ values for 3000 $\rm fb^{-1}$ integrated luminosity, and a linear fit is performed (see Figures \ref{fig:asymmetries_wo_DNN} and \ref{fig:asymmetries_DNN}). For each fit we obtained the slope value and its uncertainty, then using t-statistics we calculated the confidence level (C.L.) for the slope. 

In Figure \ref{fig:asymmetries_wo_DNN}, asymmetries are calculated without any ML discrimination whereas in Figure \ref{fig:asymmetries_DNN} a DNN classification is employed. The effect of DNN classification shows itself in the ``expected asymmetry" and ``CEDM-only asymmetry" fits. When DNN is employed the expected fit approaches to CEDM-only fit (i.e. the fit line moves from $2\sigma$ band to $1\sigma$ band as expected). Although the asymmetry values are slightly smaller than those reported in \cite{Hayreter2016} mostly because of b-tagging efficiency in detector simulation, we conclude that ML approach clearly improves the asymmetry searches.

In relevant studies asymmetries were mostly calculated using signal only events at the  parton level where background contribution were simulated by a dilution factor.  It is  anticipated that the sensitivity to triple product asymmetries get worsen at the detector level analysis. In this work, simulating the more realistic scenario with  background contamination, we applied ML techniques to filter out the signal events and showed that asymmetry calculation can be improved.

\end{document}